\documentclass[a4paper,11pt]{article}
\usepackage{jheppub}
\usepackage{amsmath,amssymb}
\usepackage{booktabs,array,graphicx,multirow}
\usepackage{xcolor}
\newcommand{\Op}[1]{\mathcal{O}_{#1}}
\newcommand{\keVnr}{\,\text{keV}_\text{nr}}
\newcommand{\tyr}{\,\text{t\,yr}}

\title{Can the ultimate direct detection experiments reveal multiple dark matter species?}
\author[a]{Carlos E.~Yaguna}
\affiliation[a]{Escuela de Física, Universidad Pedagógica y Tecnológica de Colombia,\\Avenida Central del Norte \# 39-115, Tunja, Colombia}
\emailAdd{carlos.yaguna@uptc.edu.co}

\abstract{Whether signals in direct detection experiments can establish that dark matter consists of more than one species has been studied for a long time, under various assumptions about the signal strength, the exposures of future experiments and the dark matter interactions. Two developments now allow a definite answer to this question. First, current limits, led by LZ, have become so stringent that they fix the maximum signal rate that any future experiment can observe. Second, the planned exposures of the ultimate xenon and argon detectors, the xenon observatory XLZD and the argon program of DarkSide-20k and ARGO, are well defined. In view of these developments, we revisit the question within the non-relativistic effective field theory of dark matter--nucleus scattering, so that the interactions need not be assumed either. We find that the main obstacle to establishing that there is more than one species is the small number of events that current limits still allow. In fact, even assuming the standard spin-independent interaction, XLZD with its largest envisaged exposure can reject a single species at an estimated $5\sigma$ only if the lightest dark matter species is below about $12$~GeV. A more general interaction for the single species removes this $5\sigma$ reach in xenon alone. Argon data from DarkSide-20k add little, but the planned exposure of ARGO may extend the reach to about $15$--$20$~GeV. Since these are the ultimate detectors of this kind, our results imply that direct detection is unlikely ever to tell, even at $3\sigma$, whether dark matter consists of one species or several when all of them are heavier than about $30$~GeV.}


\begin{document}
\maketitle

\section{Introduction}
\label{sec:intro}
Dark matter (DM), which accounts for about a quarter of the energy density of the Universe~\cite{Planck:2018vyg}, need not consist of a single species. All the evidence for DM is gravitational, and gravity responds only to the total DM density, so it cannot tell whether DM consists of one species or several~\cite{Zurek:2008qg}. The number of DM species, a basic property of the dark sector, must therefore be determined by experiments that probe individual DM particles, such as direct detection experiments.

Several species are, moreover, a natural possibility. Visible matter consists of several stable species, and so does DM in many particle physics models~\cite{Boehm:2003ha,Cao:2007fy,Hur:2007ur,Feldman:2010wy,Batell:2010bp,Aoki:2012ub,Bhattacharya:2013hva,Esch:2014jpa,Bernal:2018aon,BasiBeneito:2022qxd}, among which those based on a single $\mathbb{Z}_N$ symmetry are particularly simple~\cite{Yaguna:2019cvp,Belanger:2020hyh,Yaguna:2021vhb,Belanger:2022esk,Yaguna:2021rds}. In such models, each species contributes to the signal of a direct detection experiment with a recoil spectrum determined by its mass and interactions.

What an experiment observes is then the sum of their individual recoil spectra, and the question is whether such a signal can be distinguished from that of a single DM species. This question has been studied repeatedly~\cite{Profumo:2009tb,Adulpravitchai:2011ei,HerreroGarcia:2017vrl,HerreroGarcia:2018qnz,Dienes:2012cf}, and signals from several species were found to be distinguishable from those of one for sufficiently different masses and large enough exposures. These studies, however, rested on three assumptions: the size of the signal, the exposures of future experiments, and the interaction of the DM, which they took to be the standard spin-independent one.

The size of the signal and the exposures no longer need to be assumed. The limits set by LZ with $4.2\tyr$~\cite{LZ:2024zvo}, complemented by those of XENONnT~\cite{XENON:2025vwd} and PandaX-4T~\cite{PandaX:2024qfu}, are so strong that they bound the event rate in any future experiment far below the rates considered in earlier studies. At the same time, the ultimate xenon and argon detectors, designed to approach the neutrino background over most of the WIMP mass range, have well-defined planned exposures: at least $200\tyr$ and possibly up to $1000\tyr$ for XLZD~\cite{XLZD:2024nsu}, $200\tyr$ for DarkSide-20k~\cite{DarkSide-20k:2017zyg,Santone:2024ds} and about $3000\tyr$ for ARGO~\cite{Seth:2026argo}. Together, the limits and the exposures fix the maximum number of events that these detectors can observe. Whether direct detection can establish that DM consists of more than one species is thus no longer a question about hypothetical experiments, but one with a definite answer for these detectors.

The interaction of the DM, finally, need not be fixed in advance. The general framework for DM--nucleus scattering in direct detection is the non-relativistic effective theory (NREFT)~\cite{Fan:2010gt,Fitzpatrick:2012ix,Anand:2013yka}, in which the standard spin-independent interaction is one of several operators built from the momentum transfer, the relative velocity and the spins. We work within this framework. Since a single target cannot, in general, determine both the mass and the interaction of a DM species~\cite{Gluscevic:2015sqa,Edwards:2018lsl,Krauss:2018}, a single species with a different interaction may then mimic the signal of two. To our knowledge, previous studies have not considered this possibility.

In this paper we determine whether XLZD, alone or together with DarkSide-20k or ARGO, can establish that a signal as large as LZ allows comes from more than one species, by testing a single species against two, the simplest alternative. We consider WIMP-like DM, that is, particles with masses above a few GeV detected through nuclear recoils. We find that the main obstacle is the small number of events that current limits still allow. Even assuming the standard spin-independent interaction, XLZD with its largest envisaged exposure can reject a single species at an estimated $5\sigma$ only if the lightest species is below about $12$~GeV. Allowing the single species a more general interaction removes this reach in xenon, although only through a strong cancellation between two of its operators. Argon data from DarkSide-20k add little, but the larger exposure of ARGO extends the reach to about $15$--$20$~GeV. Since these are the ultimate detectors of this kind, direct detection is unlikely ever to tell, even at $3\sigma$, whether DM consists of one species or several when all of them are heavier than about $30$~GeV.

The paper is organized as follows. Section~\ref{sec:formalism} presents the direct detection formalism, section~\ref{sec:framework} defines the hypotheses, section~\ref{sec:setup} describes the experimental setup, the backgrounds and the event budget, and section~\ref{sec:stats} the statistical method. Section~\ref{sec:shape} studies the degeneracy of the spectral shapes without backgrounds, and section~\ref{sec:results} presents the results with backgrounds, the role of argon, their robustness and the case of two species with different interactions. We discuss our assumptions in section~\ref{sec:discussion} and conclude in section~\ref{sec:conclusions}. Appendix~\ref{app:calib} gives details on the calibration of the test statistic, and appendix~\ref{app:pico} shows that bubble chamber limits do not exclude the single-species fits that rely on spin-dependent operators.

\section{Direct detection formalism}
\label{sec:formalism}
The NREFT~\cite{Fan:2010gt,Fitzpatrick:2012ix,Anand:2013yka} describes DM--nucleus scattering independently of the underlying particle physics model: the interaction between a DM particle and a nucleon is written as a sum of operators built from the momentum transfer $\mathbf q$, the transverse relative velocity $\mathbf v^\perp$ and the spins of the two particles, each with an unknown coefficient. For spin-zero DM, the standard basis contains four operators, all of at most second order in $\mathbf q$ and $\mathbf v^\perp$; spin-one-half DM has these four and ten more, which involve its spin~\cite{Fitzpatrick:2012ix,Anand:2013yka,Dent:2015zpa}. To make our results independent of the DM spin, we restrict the analysis to the four operators that do not involve it. The additional operators of DM with spin give a single species more freedom to imitate two species, so our results are, if anything, optimistic where they depend on the interaction. The main limitation we find comes from kinematics and from the event budget allowed by current limits, and does not depend on the spin.

\subsection{Operators}
\label{sec:nreft}
The four operators of the standard basis that do not involve the DM spin are (with $\mathbf q\cdot\mathbf v^\perp=0$)~\cite{Fitzpatrick:2012ix,Anand:2013yka,Dent:2015zpa}
\begin{equation}
\Op1=\mathbf 1_\chi\mathbf 1_N,\qquad
\Op3=i\,\mathbf S_N\cdot\Big(\frac{\mathbf q}{m_N}\times\mathbf v^\perp\Big),\qquad
\Op7=\mathbf S_N\cdot\mathbf v^\perp,\qquad
\Op{10}=i\,\mathbf S_N\cdot\frac{\mathbf q}{m_N},
\end{equation}
where $\mathbf S_N$ is the nucleon spin and $m_N$ the nucleon mass. We take isoscalar couplings, $c_j^p=c_j^n\equiv c_j$, and write $\mathbf c=(c_1,c_3,c_7,c_{10})$. The operators couple to different nuclear responses~\cite{Fitzpatrick:2012ix}: $\Op1$ to the coherent response $W_M$; $\Op3$ to the spin-orbit response $W_{\Phi''}$, suppressed by $q^4/m_N^4$, and to the transverse spin response $W_{\Sigma'}$; $\Op7$ to $W_{\Sigma'}$, suppressed by $v_\perp^2$; and $\Op{10}$ to the longitudinal spin response $W_{\Sigma''}$, suppressed by $q^2/m_N^2$. For comparable coefficients, these factors are of order $10^{-3}$--$10^{-2}$ for $\Op{10}$ and $10^{-6}$ for $\Op7$, and $\Op3$, suppressed by $q^2v_\perp^2/m_N^2$ or $q^4/m_N^4$, is the most suppressed of the four. The spin responses require a nucleus with non-zero spin, which in natural xenon means $^{129}$Xe and $^{131}$Xe. The only interference among these operators in the elastic rate is between $\Op1$ and $\Op3$, through $W_{M\Phi''}$~\cite{Fitzpatrick:2012ix,Anand:2013yka}.

\begin{figure}[t]
\centering
\includegraphics[width=\textwidth]{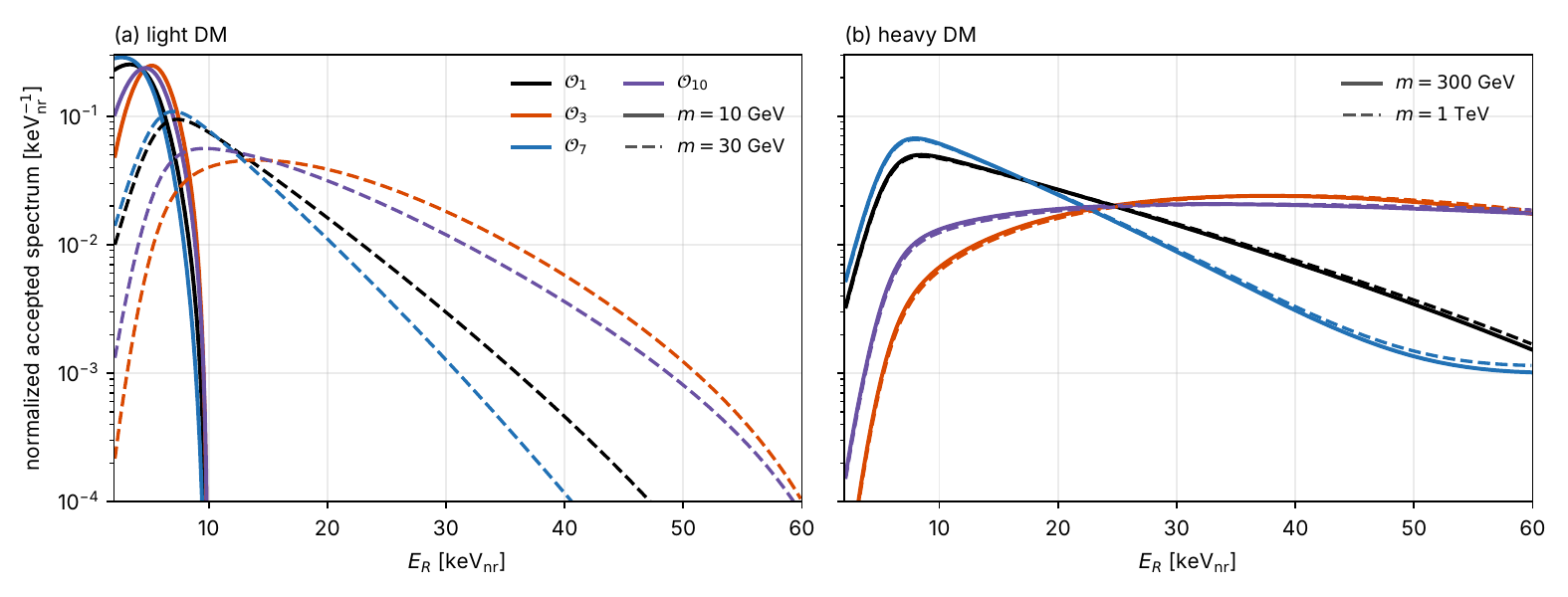}
\caption{Normalized recoil spectra in xenon for a single species with each of the operators $\Op1$, $\Op3$, $\Op7$ and $\Op{10}$ (colors), including the acceptance of the detector model of section~\ref{sec:setup} but not its energy resolution. (a) Light DM, $m=10$~GeV (solid) and $30$~GeV (dashed). (b) Heavy DM, $m=300$~GeV (solid) and $1$~TeV (dashed).}
\label{fig:ops}
\end{figure}

These responses give the operators different spectral shapes, but how different depends on the DM mass (figure~\ref{fig:ops}, in our xenon detector model of section~\ref{sec:setup}). For light DM, kinematics cuts off all the spectra at low recoil energies: at $10$~GeV they look alike, while at $30$~GeV they extend to much higher energies and differ between operators. For heavy DM, the momentum-dependent operators $\Op3$ and $\Op{10}$ are suppressed at low energies and extend to much higher ones, while the velocity-dependent $\Op7$ gives a spectrum slightly softer than that of $\Op1$. Once the DM mass is much larger than the mass of the nucleus, the reduced mass approaches $m_T$ and the shapes no longer depend on the DM mass: the spectra for $300$~GeV and $1$~TeV are almost identical, although they remain different for each operator. These features determine how well a single species with a different interaction can reproduce the spectrum of two species.

\subsection{Recoil rates}
\label{sec:rates}
For a DM particle of mass $m$ scattering on a target with isotopes $T$, mass fractions $\xi_T$ and nuclear masses $m_T$,
\begin{equation}
\frac{dR}{dE_R}=\sum_T\frac{\xi_T\,\rho}{m\,m_T}\int_{v\ge v_{\min,T}}d^3v\,f_{\rm lab}(\mathbf v)\,v\,\frac{d\sigma_T}{dE_R},\qquad
v_{\min,T}=\sqrt{\frac{m_TE_R}{2\mu_T^2}},
\end{equation}
with $\mu_T$ the DM--nucleus reduced mass. The differential cross section is a sum of nuclear responses $W_k^T(q^2)$ multiplied by DM responses that are quadratic in the coefficients $c_j$ and depend on $q^2/m_N^2$ and $v_\perp^2$~\cite{Fitzpatrick:2012ix,Anand:2013yka}. We use the shell-model responses of refs.~\cite{Fitzpatrick:2012ix,Anand:2013yka}, as tabulated in the code \texttt{WIMpy\_NREFT}~\cite{WIMpy}, for all xenon isotopes and for $^{40}$Ar, which makes up $99.6\%$ of natural argon and has spin zero, so that $\Op7$ and $\Op{10}$ do not contribute in argon. For the $\Op1$ interaction, our xenon rate agrees with the standard calculation with a Helm form factor~\cite{Helm:1956zz,Lewin:1995rx} to $1$--$2\%$ in normalization.

We use the Standard Halo Model with $\rho_0=0.3$~GeV\,cm$^{-3}$, $v_0=238$~km\,s$^{-1}$ and $v_{\rm esc}=544$~km\,s$^{-1}$, as recommended in ref.~\cite{Baxter:2021pqo}, with the annually averaged velocity of the Earth, and the same velocity distribution for all DM species.

Since the rate is quadratic in $\mathbf c$, we write it as $dR/dE_R=\mathbf c^T\,[d\mathcal M(m)/dE_R]\,\mathbf c$, where the symmetric matrix $d\mathcal M/dE_R$ contains the isotope sum, the nuclear and DM responses and the velocity integrals. The expected number of events in bin $k$ of a detector with response kernel $K_k(E_R)$ and exposure $\mathcal E$ can be written as
\begin{equation}
s_k(m,\mathbf c)=\mathbf c^T M_k(m)\,\mathbf c ,\qquad
M_k(m)=\mathcal E\int dE_R\,K_k(E_R)\,\frac{d\mathcal M(m)}{dE_R},
\label{eq:quadform}
\end{equation}
where the $4\times4$ matrices $M_k$, which include the detector response, are computed once for each mass, so profiling over $\mathbf c$ is cheap.

\section{Hypotheses}
\label{sec:framework}
Suppose that a signal is observed in a future experiment. The first hypothesis to test is a single DM species with the standard spin-independent interaction. If it is rejected, the explanations include a single species with a different interaction and more than one species, but also departures from our other assumptions, such as a different velocity distribution, inelastic scattering or light mediators, which we do not consider. Within these assumptions, establishing that DM consists of more than one species requires excluding not only a single species with the standard interaction, but also a single species with a different interaction, as described by the NREFT. We thus compare a two-component hypothesis $H_2$ (section~\ref{sec:H2}) with a set of one-component hypotheses $H_1$ of increasing freedom (section~\ref{sec:H1}): rejecting the first of them is the first step, and rejecting the others the second.

A two-component hypothesis is enough for our question: rejecting the single species is what establishes that there is more than one, and two species are the simplest alternative to it; a signal from three or more species would, in general, also be better described by two species than by one. Determining how many species there are, beyond more than one, is out of reach with the few hundred events that current limits still allow. Moreover, once a species is much heavier than the target nucleus, its spectral shape no longer depends on its mass, so that any number of such species looks like a single one, and our conclusions for heavy species hold whatever their number. Conversely, a single species that fits the signal does not exclude further species whose contributions are small or spectrally similar to those of the others: the test can only establish a lower bound on the number of species.

\subsection{The two-component hypothesis}
\label{sec:H2}
The signal hypothesis $H_2$ contains two stable species $\chi_1$ and $\chi_2$, with masses $m_1<m_2$ and local densities $\rho_i=f_i\rho_0$. In our baseline, both have the standard spin-independent interaction,
\begin{equation}
\mathcal L_{\rm NR}\supset c^{(i)}\,\Op1 ,\qquad i=1,2 .
\end{equation}
This is the natural starting point, independently of the model that produces the two species. $\Op1$ is the leading operator: it is neither momentum- nor velocity-suppressed, and it is coherently enhanced by the number of nucleons in the nucleus. It arises at the lowest dimension from the exchange of a scalar, including the Higgs boson, or of a vector, while the other operators require specific structures, such as pseudoscalar mediators or axial-vector currents, or a symmetry that removes $\Op1$. Experimental results, including the limits that fix our event budget, are also reported for this interaction. In section~\ref{sec:mixed} we also consider two species with different interactions, each with one of the operators $\Op1$, $\Op7$ and $\Op{10}$.

The two species are distinct, so their rates add incoherently,
\begin{equation}
\frac{dR_2}{dE_R}=\frac{dR_{\chi_1}}{dE_R}+\frac{dR_{\chi_2}}{dE_R}.
\end{equation}
Direct detection is only sensitive to the products of the density fractions and the scattering strengths, $f_i\,(c^{(i)})^2$, so we do not need to specify how the two species are produced. We parametrize the relative strength of the two species by
\begin{equation}
w=\frac{N_1}{N_1+N_2},
\end{equation}
the fraction of the events that the lighter species produces in the existing LZ analysis, as modeled in section~\ref{sec:setup}. The value of $w$ is model dependent; in several $\mathbb{Z}_N$ models the two species can have comparable direct detection rates~\cite{Belanger:2020hyh,Yaguna:2021vhb,Belanger:2022esk}, in which case $w$ is close to $1/2$. Because the acceptance depends on the recoil energy, the corresponding fraction in an experiment with a different response, such as argon, is in general different; we keep the physical strengths fixed when predicting other experiments, and $w$ always refers to the LZ response.

\subsection{The one-component hypotheses}
\label{sec:H1}
The alternative hypothesis $H_1$ contains a single species of mass $m$, whose interaction is a combination of the operators of section~\ref{sec:nreft} with coefficients $\mathbf c$. We consider three one-component hypotheses of increasing freedom:
\begin{enumerate}
\item $\Op1$ only, the hypothesis usually compared with two species~\cite{Profumo:2009tb,Adulpravitchai:2011ei,HerreroGarcia:2017vrl} and the first to be tested if a signal is observed;
\item the non-interfering operators $\{\Op1,\Op7,\Op{10}\}$, with arbitrary coefficients;
\item the free NREFT: $\{\Op1,\Op3,\Op7,\Op{10}\}$ with arbitrary coefficients and relative signs.
\end{enumerate}
The distinction between hypotheses 2 and 3 is physical. Since $\Op7$ and $\Op{10}$ do not interfere with $\Op1$, the rate in hypothesis 2 is a sum of spectral shapes with non-negative coefficients: a single species can only add events to the $\Op1$ spectrum. In hypothesis 3, the interference between $\Op1$ and $\Op3$ also allows it to remove events from part of the spectrum, at the price of a large coefficient for the most suppressed operator. Hypothesis 3 is the most conservative: it makes no assumption about the coefficients of the four operators. We take hypothesis 2 as our reference, since it is the most general one in which the additional operators can only add events, and use hypothesis 3 to show how much interference adds.

To identify which operators are responsible for the effect of the interaction, we also consider in section~\ref{sec:shape}, without backgrounds, each operator added separately to $\Op1$. Together with hypotheses 2 and 3, these show whether a single operator, or only a combination of them, allows a single species to imitate two.

For the second step to be meaningful, the one-component hypothesis must contain the operators of the two-component signal: if one of the species interacts through $\Op{10}$, for instance, rejecting a single $\Op1$ species tests the interaction rather than the number of species. When the two species have different interactions (section~\ref{sec:mixed}) we therefore consider only hypotheses 2 and 3 as alternatives.

\section{Experimental setup and event budget}
\label{sec:setup}
Our aim is to understand which features of the problem control the result rather than to reproduce a particular experiment, so we use simple detector models with the main features of current and planned experiments.

\subsection{Detectors}
\label{sec:detectors}
We model the response of each detector as a function of the nuclear recoil energy, with the parameters listed in table~\ref{tab:detectors}. For xenon, the nuclear-recoil acceptance is $\epsilon(E_R)=0.9/[1+e^{-(E_R-E_{\rm th})/\Delta}]$, with $E_{\rm th}=5.5\keVnr$ and $\Delta=1.1\keVnr$, and the energy resolution is Gaussian, with $\sigma_E=0.6\sqrt{E_R/\text{keV}}\,\text{keV}+0.03\,E_R$. We use this LZ-like response both for the existing LZ analysis and for XLZD, the next-generation xenon observatory, which plans an exposure of at least $200\tyr$ and possibly up to $1000\tyr$~\cite{XLZD:2024nsu}. Unless stated otherwise we take $\mathcal E_{\rm Xe}=1000\tyr$, the most favorable case; in section~\ref{sec:robust} we also lower its threshold. For argon we follow DarkSide-20k, now under construction at the Gran Sasso laboratory~\cite{DarkSide-20k:2017zyg,Santone:2024ds,DarkSide-20k:2026daq}, with an exposure of $200\tyr$, and model ARGO, a detector with a fiducial mass of $300$~t that is at the conceptual stage, with the same detector response and its planned exposure of $3000\tyr$~\cite{Seth:2026argo}. The argon threshold is high enough that a DM particle lighter than about $15$~GeV produces almost no events.

\begin{table}[ht]
\centering
\small
\begin{tabular}{@{}>{\raggedright\arraybackslash}p{0.17\textwidth}>{\raggedright\arraybackslash}p{0.4\textwidth}>{\raggedright\arraybackslash}p{0.35\textwidth}@{}}
\toprule
 & xenon (LZ, XLZD) & argon (DarkSide-20k, ARGO) \\
\midrule
energy window & $2$--$60\keVnr$, $1\keVnr$ bins & $30$--$200\keVnr$, $10\keVnr$ bins \\
acceptance & sigmoid, maximum $0.9$, half maximum at $5.5\keVnr$ & flat, $0.7$ \\
resolution & $0.6\sqrt{E_R/\text{keV}}\,\text{keV}+0.03\,E_R$ & $10\%$ \\
exposure & $4.2\tyr$ (LZ); $200$ or $1000\tyr$ (XLZD) & $200\tyr$ (DarkSide-20k); $3000\tyr$ (ARGO) \\
neutrinos & $0.57$ per t\,yr & $0.012$ per t\,yr \\
other & electron-recoil leakage, $0.15$ per t\,yr ($\pm20\%$) & instrumental, $0.1$ per $200\tyr$ ($\pm50\%$) \\
\bottomrule
\end{tabular}
\caption{Detector models. Background rates are the accepted events in the recoil energy window; the neutrino background is dominated by $^8$B in xenon and by atmospheric neutrinos in argon.}
\label{tab:detectors}
\end{table}

\subsection{Backgrounds}
\label{sec:bkg}
The neutrino background limits the reach of future experiments~\cite{Billard:2013qya,OHare:2021utq}, and XENONnT, PandaX-4T and LZ have reported first evidence of nuclear recoils from solar $^8$B neutrinos~\cite{XENON:2024ijk,PandaX:2024muv,LZ:2025light}. We include coherent elastic neutrino--nucleus scattering~\cite{Freedman:1973yd} of $^8$B, \emph{hep}, atmospheric and diffuse supernova neutrinos, with the spectra of ref.~\cite{OHare:2016pjy}, the $^8$B flux normalized to the SNO measurement~\cite{SNO:2011hxd}, and normalization uncertainties of $4\%$, $30\%$, $20\%$ and $50\%$, respectively. Table~\ref{tab:detectors} gives the resulting rates; in xenon, almost all $^8$B events lie below $5\keVnr$. We model electron recoils that leak into the nuclear-recoil region as a flat spectrum; since this is the most uncertain ingredient, we also consider rates of $0$ and $1.5$ events per $\text{t\,yr}$, besides the reference value. In argon, pulse-shape discrimination makes electron-recoil leakage negligible, and we include only a small instrumental background besides the neutrinos.

\subsection{Event budget}
\label{sec:budget}
The future signal cannot exceed what the LZ analysis of $4.2\tyr$~\cite{LZ:2024zvo} allows. In our LZ-like response, the strongest published $90\%$ CL limit on the spin-independent cross section, $2.2\times10^{-48}$~cm$^2$ at $40$~GeV~\cite{LZ:2024zvo}, corresponds to $N_{\rm cap}\simeq1.8$ accepted signal events. At light masses, this cap is consistent with the more recent LZ search for light DM with $5.7\tyr$~\cite{LZ:2025light}: at $9$~GeV it corresponds to $\sigma_p\simeq1.0\times10^{-46}$~cm$^2$, while that search excludes $\sigma_p>1.1\times10^{-46}$~cm$^2$. We use $N_{\rm cap}$ as the maximum number of accepted events in LZ for any two-component signal. The cap applies to the sum of the two species, not to each of them separately: we distribute it between them according to $w$, whatever the interaction of each, and fix the physical strengths $f_i(c^{(i)})^2$ accordingly. The same strengths then determine the signal in every future experiment, including argon. If $Y_{ai}$ is the number of accepted events of species $i$ in experiment $a$ per unit strength, the expected number of signal events in experiment $a$ is
\begin{equation}
N_a=N_{\rm cap}\left[w\,\frac{Y_{a1}}{Y_{{\rm LZ},1}}+(1-w)\,\frac{Y_{a2}}{Y_{{\rm LZ},2}}\right].
\label{eq:Na}
\end{equation}
In the future xenon experiment, which has the same response as LZ, both ratios are equal to the ratio of the exposures, $\mathcal E_{\rm Xe}/4.2\tyr$, so the maximal signal is about $420$ events for $1000\tyr$ and $85$ for $200\tyr$; in argon, the ratios depend on the masses and on the interactions.

For comparison, the benchmark cross section of refs.~\cite{HerreroGarcia:2017vrl,HerreroGarcia:2018qnz}, $\sigma_p=10^{-45}$~cm$^2$, corresponds to between about $20$ and $460$ times more events in LZ than this cap for masses between $10$ and $200$~GeV, and that of ref.~\cite{Profumo:2009tb}, $10^{-44}$~cm$^2$, to ten times more still.

Treating the cap as a fixed number of events ignores the dependence of the LZ limit on the shape of the spectrum, but, within our estimate, it is conservative. To estimate the effect, we computed the median $90\%$ CL limit, in number of events, for $4.2\tyr$ in our LZ-like response with the backgrounds of section~\ref{sec:bkg}. For all the two-component spectra we consider, including two species with different interactions and $w$ between $0.25$ and $0.75$, it is at least as large as that of a $40$~GeV $\Op1$ species, and larger by up to about $45\%$ for the lightest species, whose events overlap with the $^8$B background; near the boundaries of the region where two species can be distinguished, the excess is $5$--$20\%$. A shape-dependent cap would therefore allow a somewhat larger signal when a light species contributes. At $(m_1,m_2)=(10,350)$~GeV, well inside that region in xenon, the signal would be $23\%$ larger and $T_A$ would increase from $52$ to $73$; at $(15,350)$~GeV, near its edge, the increase is $8\%$ and $T_A$ changes from $17$ to $19$; with ARGO, $T_A$ at $(20,350)$~GeV changes from $30$ to $32$. Within our model, the boundary of the region moves by about $1$~GeV towards larger masses. These are median expected limits, whereas the cap is derived from the observed one; against an observed limit, we have checked the cap only at $9$~GeV, and a full recast of the LZ likelihood could change it by more. Unless stated otherwise, the signal saturates the fixed cap $N_{\rm cap}$.

\section{Statistical method}
\label{sec:stats}
We compare the hypotheses with a binned Poisson likelihood,
\begin{equation}
-2\ln\mathcal L=2\sum_{a,k}\Big[\mu_{ak}-n_{ak}+n_{ak}\ln\frac{n_{ak}}{\mu_{ak}}\Big]+\sum_j\Big(\frac{\theta_j-1}{\sigma_j}\Big)^2,
\qquad
\mu_{ak}=s_{ak}+\sum_j\theta_j\,b_{ak}^{(j)} ,
\end{equation}
where $a$ labels the experiments, $b^{(j)}$ are the background templates with normalizations $\theta_j$ and uncertainties $\sigma_j$, and the signal $s_{ak}$ is given by eq.~\eqref{eq:quadform}. The test statistic is
\begin{equation}
T=-2\ln\frac{\max_{H_1}\mathcal L}{\max_{H_2}\mathcal L},
\end{equation}
where the maximization over $H_1$ includes $m$, the coefficients $\mathbf c$ and the background normalizations, and that over $H_2$ the two masses, the two strengths and the background normalizations. The signal normalization is free in both hypotheses; in the combined analysis of xenon and argon, the same couplings and mass determine the signal in both targets. For the median sensitivity we evaluate $T$ on the Asimov data set~\cite{Cowan:2010js}, with $n_{ak}$ equal to the expected counts for the two-component signal being tested, and call the result $T_A$.

To separate the information carried by the spectral shapes from the effect of backgrounds, we also consider the case without backgrounds. With a free signal normalization, $T_A$ then depends only on the normalized accepted spectra and on the number of signal events. For a two-component spectrum $p_2$ we compute
\begin{equation}
D_{\min}=\min_{m,\hat{\mathbf c}}D_{\rm KL}\big(p_2\,\|\,p_1(m,\hat{\mathbf c})\big),
\end{equation}
where $D_{\rm KL}(p\|q)=\sum_kp_k\ln(p_k/q_k)$ is the Kullback--Leibler divergence between the accepted spectra in the $58$ bins of $1\keVnr$ between $2$ and $60\keVnr$, each normalized to unit sum, and $\hat{\mathbf c}$ the direction of the coefficient vector. For $N_s$ signal events, $T_A=2N_sD_{\min}$, so that $N_s=T_A/(2D_{\min})$ events are needed to reach a given $T_A$. We minimize over $m$ on a logarithmic grid from $6$~GeV to $30$~TeV, without restricting it to the interval between $m_1$ and $m_2$, and refine the minimum, with several starting directions and both relative signs of $c_1$ and $c_3$. With backgrounds, we evaluate $T_A$ directly from the likelihood: they compete with the signal where the spectrum carries the information, and their normalizations are uncertain.

To convert $T$ into a significance, we need its distribution under $H_1$. For the $\Op1$ hypothesis, $H_1$ is the limit of $H_2$ in which one species vanishes, but this limit lies on the boundary of the parameter space of $H_2$, where the mass of the vanishing species is not identified. For the other two hypotheses, $H_1$ contains operators that $H_2$ lacks, and the two hypotheses are not nested. In neither case does the distribution of $T$ follow Wilks' theorem, and we calibrate it with pseudo-experiments for five benchmarks that cover the three one-component hypotheses (appendix~\ref{app:calib}). We find that a $3\sigma$ ($5\sigma$) rejection of $H_1$ requires $T\simeq14$ ($T\simeq31$), rather than the values $9$ ($25$) suggested by a $\chi^2$ with one degree of freedom, and we use $T_A=14$ and $T_A=31$ as the $3\sigma$ and $5\sigma$ thresholds for the median experiment.

For the non-interfering hypothesis, $T$ is zero or slightly negative in $13$--$39\%$ of the pseudo-experiments generated under the one-component hypothesis, and otherwise follows approximately a $\chi^2$ distribution with $2.1$--$3$ degrees of freedom; under the two-component hypothesis, the median of $T$ agrees with $T_A$ to within $3\%$, so that $T_A$ is a reliable estimate of the median sensitivity. Across the five benchmarks, the fitted thresholds range from $11.6$ to $15.3$ (from $29.1$ to $32.9$). At one benchmark we generated $3000$ pseudo-experiments under the one-component hypothesis: none of them gives $T>14$, and the empirical $3\sigma$ quantile is about $12$, so the $3\sigma$ threshold is conservative. The $5\sigma$ threshold, at $p\simeq3\times10^{-7}$, relies on extrapolating the fitted tail and is therefore an estimate; the $3\sigma$ reaches, which are calibrated directly, lead to the same conclusions.

The same thresholds can be used for the other two hypotheses, which we checked at one benchmark each. For a single $\Op1$ species they are the same. For the free NREFT, which is more flexible than the two-component hypothesis and gives $T\le0$ in most pseudo-experiments, they are slightly lower, $T\simeq11.6$ ($29$), so our thresholds are conservative for it. For both hypotheses, the median of $T$ under the two-component hypothesis is about $11\%$ below $T_A$, which does not change our conclusions.

Our approach follows ref.~\cite{HerreroGarcia:2017vrl}, which compared one and two species through the difference of the best-fit $\chi^2$ of the two hypotheses and its Asimov value; ref.~\cite{Profumo:2009tb} instead required that a $\chi^2$ goodness-of-fit test exclude the one-component hypothesis at $95\%$ CL. Our analysis differs from ref.~\cite{HerreroGarcia:2017vrl} in three respects: the likelihood is Poisson rather than Gaussian, backgrounds are included with profiled normalizations, and the significance is calibrated with pseudo-experiments rather than obtained from a $\chi^2$ distribution with as many degrees of freedom as the difference in the number of parameters.

\section{Spectral degeneracy without backgrounds}
\label{sec:shape}
\begin{figure}[t]
\centering
\includegraphics[width=\textwidth]{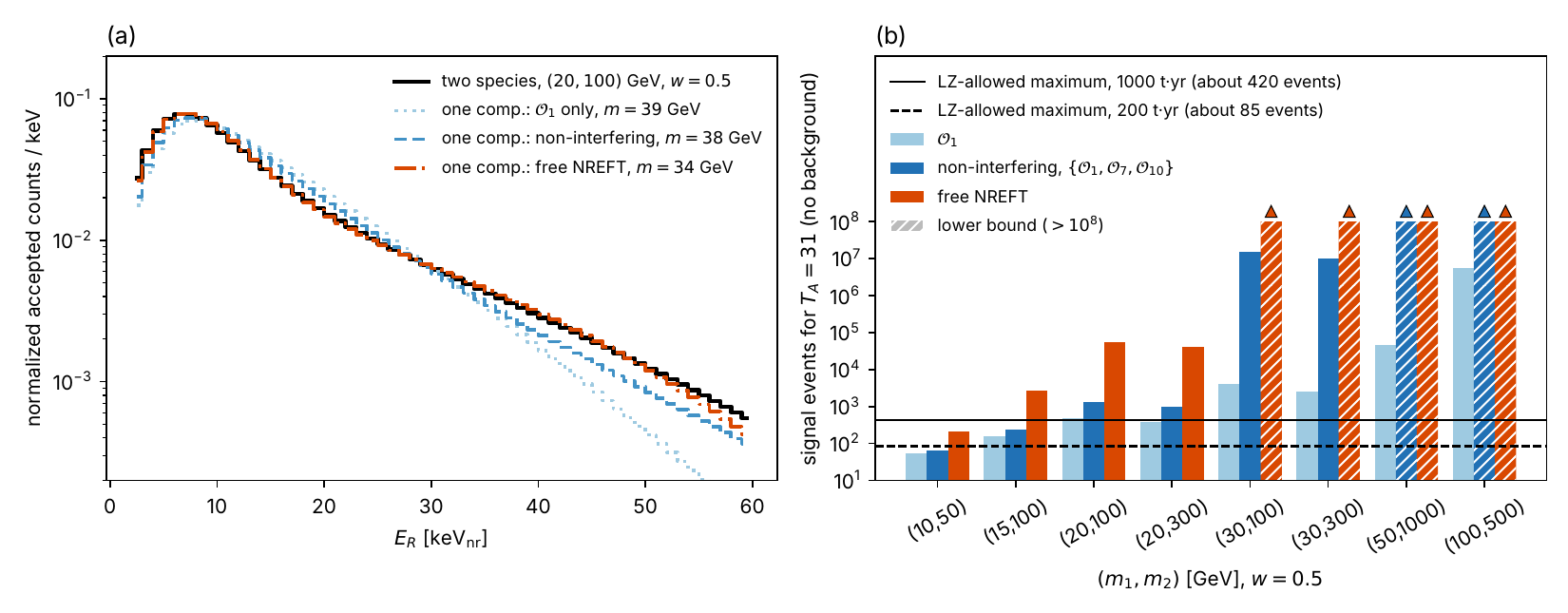}
\caption{Spectral degeneracy without backgrounds in xenon. (a) The two-component spectrum for $(m_1,m_2)=(20,100)$~GeV and $w=0.5$, and its best imitations by a single species with $\Op1$ only, with the non-interfering operators, and with the free NREFT. (b) Number of signal events needed to reach the $5\sigma$ threshold, $T_A=31$, in the absence of backgrounds, $15.5/D_{\min}$, for the three one-component hypotheses, compared with the maximal number of events allowed by LZ in $200$ and $1000\tyr$. Hatched bars are lower bounds. Since backgrounds can only increase these numbers, a benchmark above the LZ-allowed maximum cannot be identified whatever the backgrounds, while one below it may still fail once backgrounds are included (section~\ref{sec:results}).}
\label{fig:shape}
\end{figure}

We begin with the information carried by the shapes of the spectra, without backgrounds, in our xenon response and for the baseline of two $\Op1$ species. Figure~\ref{fig:shape}a shows the two-component spectrum for $m_1=20$~GeV, $m_2=100$~GeV and $w=0.5$, together with its best imitations, that is, the best-fitting single species, in three one-component hypotheses. The single $\Op1$ species, with $m\simeq39$~GeV, fails to reproduce the hard tail above $30\keVnr$ that the heavy species provides. The non-interfering hypothesis comes closer by adding the spin-dependent $\Op7$ and $\Op{10}$ shapes, and the free NREFT reproduces the spectrum almost exactly.

Figure~\ref{fig:shape}b and table~\ref{tab:shape} quantify this through the number of events needed to reach the $5\sigma$ threshold, $T_A=31$. Three features stand out.

\begin{table}[!b]
\centering
\small
\begin{tabular}{@{}lrrrrrr@{}}
\toprule
 & & \multicolumn{3}{c}{$\Op1$ plus one operator} & & \\
\cmidrule(lr){3-5}
$(m_1,m_2)$ [GeV] & $\Op1$ & $+\Op{10}$ & $+\Op7$ & $+\Op3$ & non-interfering & free NREFT \\
\midrule
$(10,50)$   & 55 & 55 & 58 & 173 & 63 & 214 \\
$(15,100)$  & 157 & 168 & 176 & 162 & 242 & 2\,700 \\
$(20,100)$  & 470 & 560 & 590 & 470 & 1\,290 & 55\,000 \\
$(20,300)$  & 370 & 450 & 460 & 370 & 980 & 40\,000 \\
$(30,100)$  & 4\,000 & 9\,400 & 9\,600 & 4\,000 & $1.5\times10^7$ & $>10^8$ \\
$(30,300)$  & 2\,500 & 5\,300 & 5\,100 & 2\,500 & $1.0\times10^7$ & $>10^8$ \\
$(50,1000)$ & 46\,000 & $2.3\times10^6$ & $8.9\times10^5$ & 58\,000 & $>10^8$ & $>10^8$ \\
$(100,500)$ & $5.6\times10^6$ & $5.9\times10^7$ & $>10^8$ & $>10^8$ & $>10^8$ & $>10^8$ \\
\bottomrule
\end{tabular}
\caption{Number of signal events needed to distinguish two species from one at $5\sigma$ ($T_A=31$), without backgrounds, in xenon, for $w=0.5$, for the three one-component hypotheses and for $\Op1$ plus each of the other operators separately. Entries $>10^8$ are lower bounds: the fit cannot resolve smaller values of $D_{\min}$. For comparison, LZ allows at most about $85$ ($420$) events in $200$ ($1000$) t\,yr.}
\label{tab:shape}
\end{table}

First, kinematics alone makes heavy species hard to distinguish. Even against a single $\Op1$ species, more than $10^3$ events are needed when $m_1\gtrsim30$~GeV, and more than $10^6$ when both masses are of order $100$~GeV or above. As the DM mass becomes much larger than the mass of the nucleus, the reduced mass approaches $m_T$ and the spectrum loses its dependence on the DM mass, so that two heavy species look like one, whatever their mass ratio. Only for $m_1\lesssim20$~GeV is the number of events needed comparable to what LZ allows. The value of about $20$~GeV is set by the event budget, while kinematics explains why the number of events needed keeps growing at higher masses. Since backgrounds can only increase the number of events needed, these numbers are lower bounds: they show where the window is closed even without backgrounds, while whether the rest of it survives depends on the backgrounds, as we show in section~\ref{sec:results}.

Second, the non-interfering operators change little for $m_1\lesssim20$~GeV. Added separately to $\Op1$, $\Op{10}$ and $\Op7$ increase the number of events needed by at most about $25\%$, and together by up to a factor of about $3$. For heavier species, instead, the spin-dependent shapes can fill in the spectrum almost exactly: for $m_1\gtrsim30$~GeV the best imitations consist mostly of $\Op7$ and $\Op{10}$, with $D_{\min}$ of order $10^{-6}$ or smaller.

Third, the free NREFT removes most of the remaining difference, and it does so through a combination of operators. For $m_1\simeq15$--$20$~GeV, the best imitation combines destructive interference between $\Op1$ and $\Op3$, which suppresses the spectrum at intermediate momentum transfer while the $\Op3$ term restores a hard tail, with an incoherent $\Op7$ contribution that fills in the low-energy part. Neither ingredient suffices alone: $\Op1$ and $\Op3$ together do not improve on $\Op1$, and adding $\Op7$ alone improves on it only slightly. The interference cancels about $80\%$ of the $\Op1$ and $\Op3$ contributions and requires, with the operators normalized as in section~\ref{sec:nreft}, $c_3/c_1\sim10^3$, a large hierarchy given that $\Op3$ is the most suppressed operator. If the destructive interference is limited to $50\%$ of the $\Op1$ and $\Op3$ contributions, the numbers of events needed for $(15,100)$ and $(20,100)$~GeV fall from $2\,700$ and $55\,000$ to $320$ and $2\,700$, within a factor of about two of the non-interfering hypothesis: the additional freedom of the free NREFT comes from strong cancellations. Only for the lightest benchmark, $(10,50)$~GeV, does adding $\Op3$ alone account for most of the effect.

\section{Results with backgrounds}
\label{sec:results}
\subsection{Xenon alone}
\label{sec:xenon}
\begin{figure}[p]
\centering
\includegraphics[width=\textwidth]{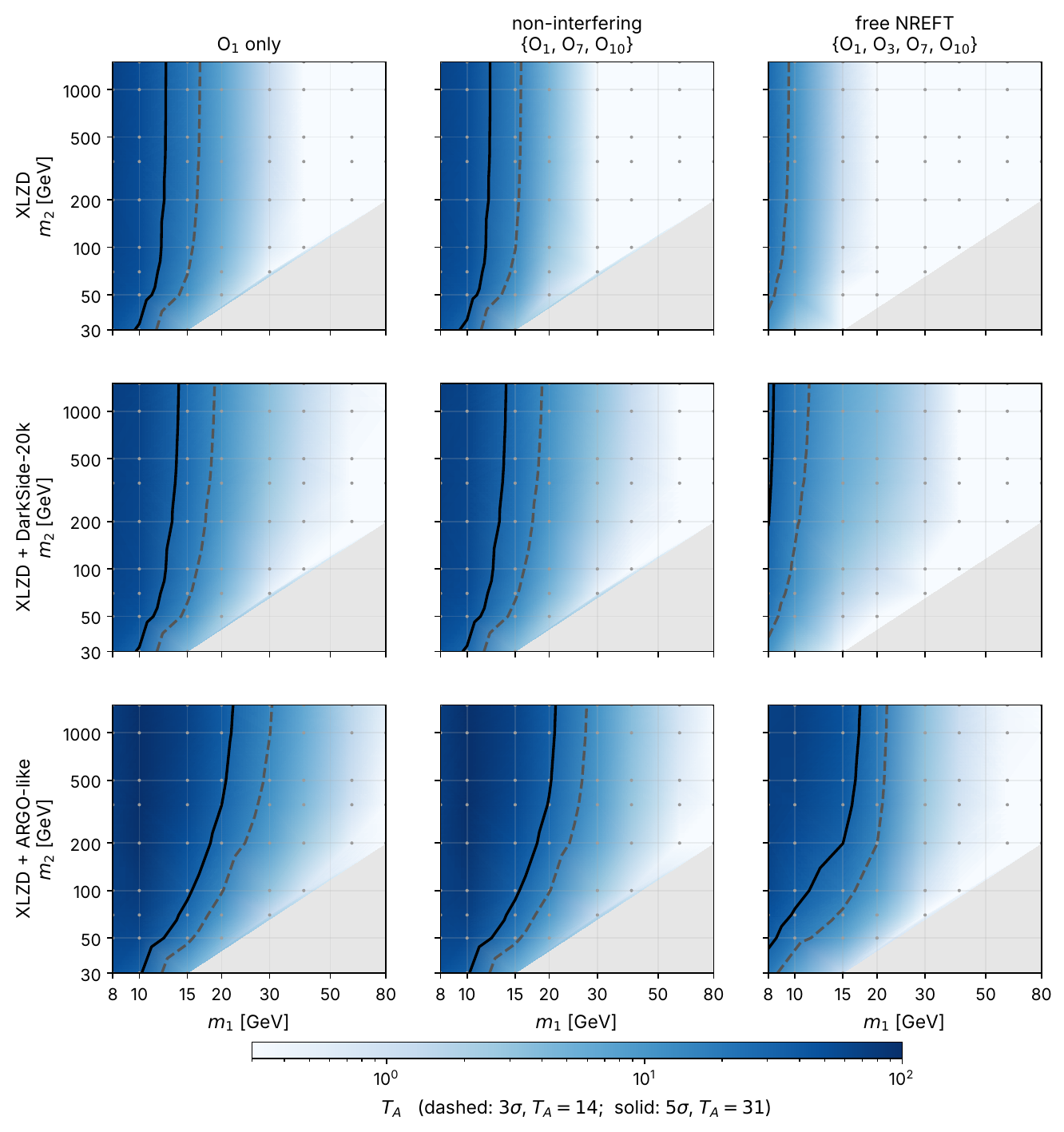}
\caption{Expected test statistic $T_A$ for rejecting one species in favor of two, as a function of the two masses, for $w=0.5$, a signal at the maximum allowed by LZ, and an electron-recoil leakage of $0.15$ events per t\,yr. Rows: XLZD ($1000\tyr$ of xenon); XLZD and DarkSide-20k ($200\tyr$ of argon); XLZD and an ARGO-like experiment ($3000\tyr$ of argon). Columns: one-component hypothesis with $\Op1$ only, with the non-interfering operators, and with the free NREFT. Dashed contours correspond to $3\sigma$ for the median experiment, as calibrated with pseudo-experiments, and solid contours to the estimated $5\sigma$, from the extrapolated tail. Grey dots show the grid points, which satisfy $m_2\ge1.8\,m_1$; the hatched region lies outside the grid.}
\label{fig:maps}
\end{figure}

We now include the backgrounds, which determine how much of the information in the spectral shapes survives. Figure~\ref{fig:maps} shows $T_A$ in the plane of the two masses, with the signal at the maximum allowed by LZ and the backgrounds of section~\ref{sec:setup}. We take $w=0.5$, for which both species contribute equally to the signal in LZ; as we show in section~\ref{sec:w}, this is close to the most favorable case, so the results of figure~\ref{fig:maps} are optimistic in this respect. The first row corresponds to XLZD with $1000\tyr$, and the three columns to the one-component hypotheses $\Op1$, non-interfering and free NREFT. In the language of section~\ref{sec:framework}, the first column corresponds to the first step, rejecting a single species with the standard interaction, and the other two to the second, excluding a single species with a different interaction.

\begin{figure}[t]
\centering
\includegraphics[width=\textwidth]{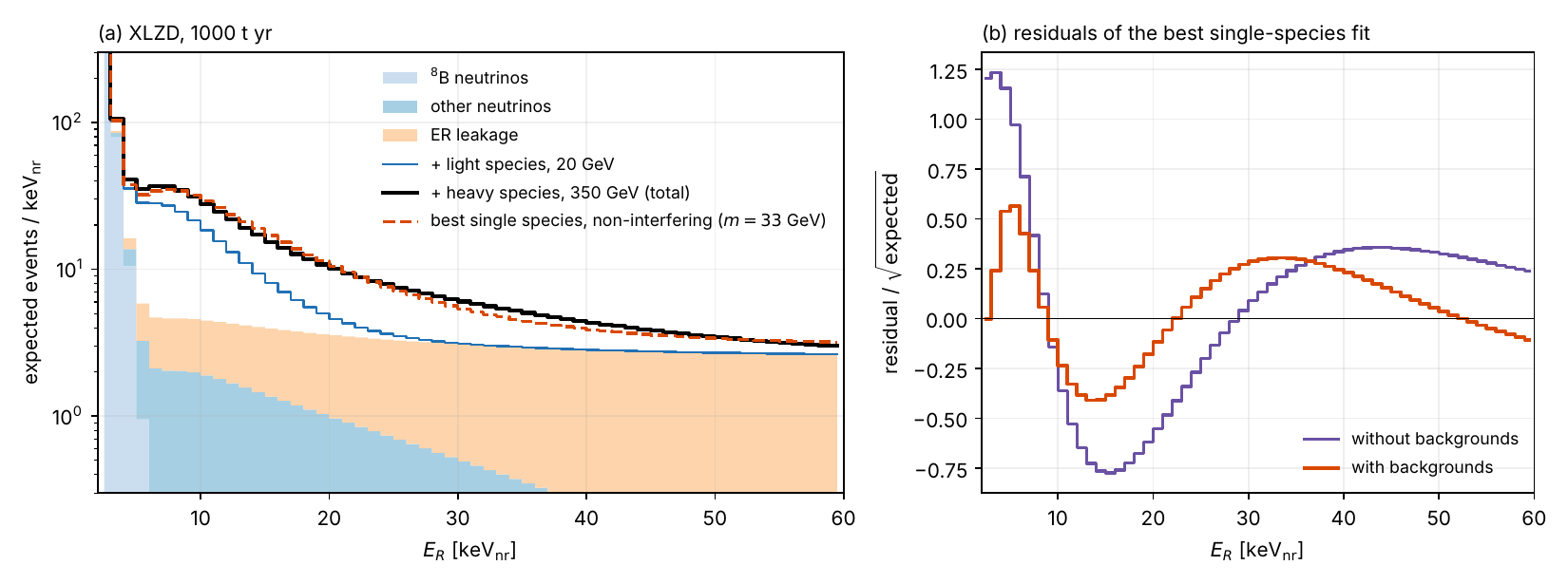}
\caption{How backgrounds remove the information, for $(m_1,m_2)=(20,350)$~GeV, $w=0.5$, XLZD with $1000\tyr$ and a signal at the maximum allowed by LZ. (a) Expected spectrum: backgrounds (stacked), the contributions of the two species, and the best fit of a single species with the non-interfering operators. (b) Residuals of the best single-species fit, in units of the square root of the expected counts per bin, with and without backgrounds.}
\label{fig:bkgspec}
\end{figure}

The window in which two species can be identified is narrow. With the non-interfering hypothesis, the median experiment reaches $5\sigma$ only for $m_1\lesssim12$~GeV and $3\sigma$ for $m_1\lesssim15$~GeV, almost independently of $m_2$. Backgrounds hide the two parts of the spectrum that carry the information: $^8$B neutrinos the steep low-energy part of the light species, and electron-recoil leakage the high-energy tail of the heavy one. Figure~\ref{fig:bkgspec} illustrates this for $m_1=20$~GeV and $m_2=350$~GeV. The best single species, with $m\simeq33$~GeV, reproduces the total spectrum to within about $15\%$ in every bin. Without backgrounds, its residuals are largest below $8\keVnr$, around $15\keVnr$ and above $35\keVnr$; with backgrounds, they are reduced everywhere, and most strongly below $8\keVnr$ and above $40\keVnr$, where $^8$B neutrinos and electron-recoil leakage dominate. As a result, $T_A=4.3$, while the same number of signal events without backgrounds would give $T_A\simeq14$.

In xenon alone, the freedom of the interaction matters only through interference. Against a single $\Op1$ species the window is almost the same as against the non-interfering hypothesis, so the limitation is the event budget rather than the interaction. Against the free NREFT, no point reaches $5\sigma$, and $3\sigma$ is reached only at the lowest mass of our grid, $m_1=8$~GeV.

With the baseline XLZD exposure of $200\tyr$, the maximal signal is about $85$ events and no point reaches $5\sigma$. For the non-interfering hypothesis and $w=0.5$, $T_A$ is $15$ at $(m_1,m_2)=(8,350)$~GeV, just above the $3\sigma$ threshold, $14$ at $(10,350)$~GeV, $13$ at $(8,50)$~GeV and $10$ at $(10,50)$~GeV, and only about $4$ for $m_1\simeq15$~GeV.

\subsection{Adding argon}
\label{sec:argon}
On its own, an argon experiment cannot test the number of species: its threshold of $30\keVnr$ makes it almost blind to a light species with $m_1\lesssim15$--$20$~GeV, the only range where the spectral information is sufficient, while for heavier species it observes too few events. Argon data are therefore useful only in combination with a xenon spectrum, and we always consider them together with XLZD, even though DarkSide-20k will start taking data earlier. The complementarity of different targets for determining the properties of a single DM species has been studied extensively~\cite{Pato:2010zk,Ruppin:2014bra,Yaguna:2019llp}. An argon target differs from xenon in two ways that matter here. Its threshold is too high for the light species, and its nucleus has no spin, so $\Op7$ and $\Op{10}$ produce no events. A combined analysis therefore tests whether the ratio of the argon and xenon rates and the argon spectrum are consistent with a single species.

DarkSide-20k does not change the picture (second row of figure~\ref{fig:maps}). With the signal capped by LZ, it records only between $2$ and $11$ events, depending on the masses. These events increase $T_A$ by a few units, and by at most about $10$ at the lightest masses, where xenon alone already has the largest sensitivity. As a result, one more point reaches $3\sigma$ against the non-interfering hypothesis, and five points at $m_1=8$~GeV reach $5\sigma$ against the free NREFT, but the window does not move.

With the exposure of ARGO, $3000\tyr$, the argon spectrum contains between $23$ and $160$ events, depending on the masses, enough to constrain the mass of the heavy species directly (third row). The $5\sigma$ region extends from $m_1\simeq15$~GeV at $m_2\simeq100$~GeV to about $20$~GeV for $m_2\gtrsim500$~GeV, and the $3\sigma$ region from about $20$~GeV to about $25$~GeV for $m_2\gtrsim300$~GeV. These values are read from the interpolated contours; on our grid, $5\sigma$ against the non-interfering hypothesis is reached at $m_1=15$~GeV for $m_2\ge100$~GeV, and at $m_1=20$~GeV only for $m_2\ge500$~GeV (table~\ref{tab:counts}). At $m_1=20$~GeV and $m_2=350$~GeV, $T_A$ increases from $4.3$ to $30$. The choice between $\Op1$ and the non-interfering hypothesis matters little, and even the free NREFT can be rejected at $5\sigma$ for $m_1\lesssim15$~GeV. For $m_1\gtrsim30$~GeV, however, the non-interfering and free-NREFT hypotheses are never rejected at $3\sigma$, and a single $\Op1$ species only marginally, for $m_2\gtrsim1$~TeV: the light species is then too heavy for the xenon spectrum to separate it from the heavy one, and the event budget is too small for the argon spectrum to help.

Table~\ref{tab:counts} summarizes the reach of each configuration, including the baseline XLZD exposure, by the largest mass of the lighter species at which the median experiment rejects each one-component hypothesis. Across the range of fitted $5\sigma$ thresholds, $T=29.1$--$32.9$, the interpolated $5\sigma$ boundary for $m_2\gtrsim500$~GeV moves by about $1$~GeV or less: from $12.4$ to $11.8$~GeV in xenon alone and from about $21$ to $20$~GeV with ARGO, against the non-interfering hypothesis. Benchmarks with $T_A$ within this range, such as $(20,350)$~GeV with ARGO, cannot be classified unambiguously, and for them we quote $T_A$ itself.

\begin{table}[t]
\centering
\begin{tabular}{@{}lccc@{}}
\toprule
 & $\Op1$ & non-interfering & free NREFT \\
\midrule
XLZD, $200\tyr$ & -- [10] & -- [10] & -- [--] \\
XLZD, $1000\tyr$ & 10 [15] & 10 [15] & -- [8] \\
XLZD + DarkSide-20k & 10 [15] & 10 [15] & 8 [10] \\
XLZD + ARGO-like & 20$^a$ [30$^c$] & 20$^b$ [20] & 15 [20] \\
\bottomrule
\end{tabular}
\caption{Largest mass of the lighter species, in GeV, among the grid values $m_1=8$, $10$, $15$, $20$, $30$, $40$, $60$, $80$~GeV, at which the median experiment rejects the one-component hypothesis at $5\sigma$ [$3\sigma$] for at least one $m_2$, for $w=0.5$; a dash means that no point does. Only for $m_2\ge350$~GeV ($^a$), $500$~GeV ($^b$) or $1$~TeV ($^c$). XLZD + DarkSide-20k and XLZD + ARGO-like assume $1000\tyr$ of xenon.}
\label{tab:counts}
\end{table}

\subsection{Relative strength of the two species}
\label{sec:w}
\begin{figure}[t]
\centering
\includegraphics[width=0.92\textwidth]{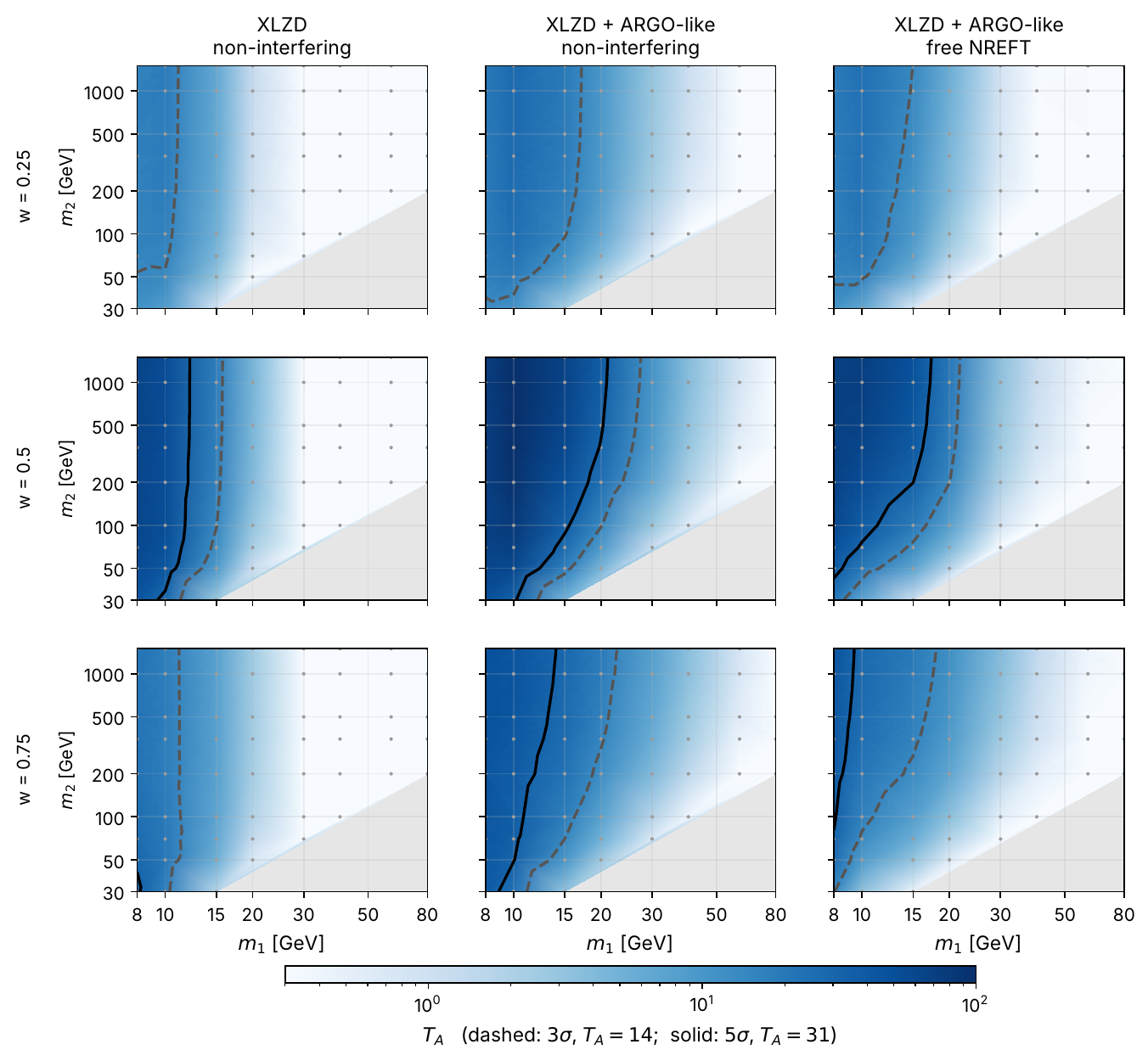}
\caption{As figure~\ref{fig:maps}, for $w=0.25$, $0.5$ and $0.75$ (rows), for xenon alone with the non-interfering hypothesis and for xenon and ARGO with the non-interfering and free-NREFT hypotheses (columns).}
\label{fig:w}
\end{figure}

So far we have taken $w=0.5$. Figure~\ref{fig:w} shows the same analysis for $w=0.25$, $0.5$ and $0.75$. The position of the boundary in $m_1$ hardly changes: it is set by kinematics and by the total event budget, which do not depend on $w$. The significance within the window, instead, depends strongly on $w$, and is largest when both species contribute comparably. For $w=0.25$, the light species is too weak to stand out above the $^8$B background, and no point reaches $5\sigma$ even with ARGO. For $w=0.75$, the heavy species is too weak for its tail to stand out above the electron-recoil leakage in xenon; ARGO recovers much of the sensitivity, because it sees the heavy species without this background.

\subsection{Robustness}
\label{sec:robust}
\begin{figure}[t]
\centering
\includegraphics[width=\textwidth]{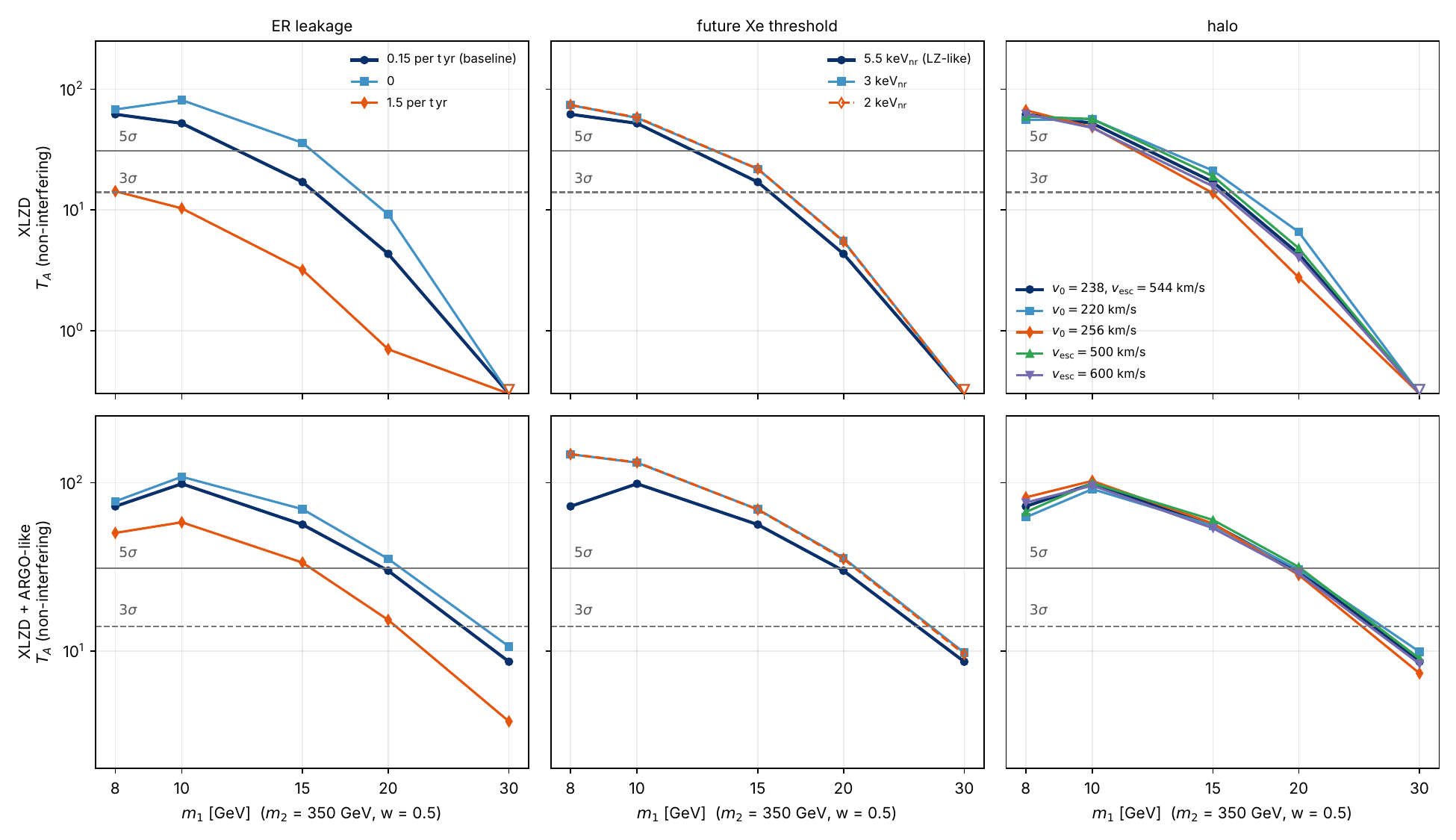}
\caption{Robustness of the results along the line $m_2=350$~GeV, for $w=0.5$ and the non-interfering hypothesis, for xenon alone (top) and xenon and ARGO (bottom). Left: electron-recoil leakage. Center: threshold of the future xenon experiment, with the signal still capped in the LZ response. Right: halo parameters. Horizontal lines show the $3\sigma$ and $5\sigma$ thresholds. The curves for thresholds of $3$ and $2\keVnr$ almost coincide. Open triangles on the lower edge mark values of $T_A$ below the plotted range.}
\label{fig:robust}
\end{figure}

Figure~\ref{fig:robust} shows how $T_A$ changes along the line $m_2=350$~GeV, for $w=0.5$ and the non-interfering hypothesis, when we vary the electron-recoil leakage, the threshold of the future xenon experiment and the halo parameters.

Of the variations we consider, the electron-recoil leakage has the largest effect. With $1.5$ events per $\text{t\,yr}$, xenon alone loses all its $5\sigma$ reach along this line and barely reaches $3\sigma$ at $m_1=8$~GeV. With ARGO, the $5\sigma$ reach survives up to $m_1\simeq15$~GeV. Argon thus makes the conclusion robust against the least known background.

A lower threshold in the future xenon experiment can increase the number of events from the light species without violating the LZ limit, which is set with the higher LZ threshold. With the acceptance shifted to $3\keVnr$, $T_A$ increases by up to a factor of two at the lowest masses in combination with ARGO, but the boundary of the window moves by only $1$--$2$~GeV. Lowering it further to $2\keVnr$ gives no additional gain in our setup, since the analysis range starts at $2\keVnr$ and the $^8$B background grows.

Varying the halo parameters within $v_0=220$--$256$~km\,s$^{-1}$ and $v_{\rm esc}=500$--$600$~km\,s$^{-1}$, consistently in the signal and in both hypotheses, changes $T_A$ by up to about $25\%$ near the $3\sigma$ and $5\sigma$ thresholds and moves the boundaries by $1$--$2$~GeV.

\subsection{Two species with different interactions}
\label{sec:mixed}
\begin{figure}[t]
\centering
\includegraphics[width=0.95\textwidth]{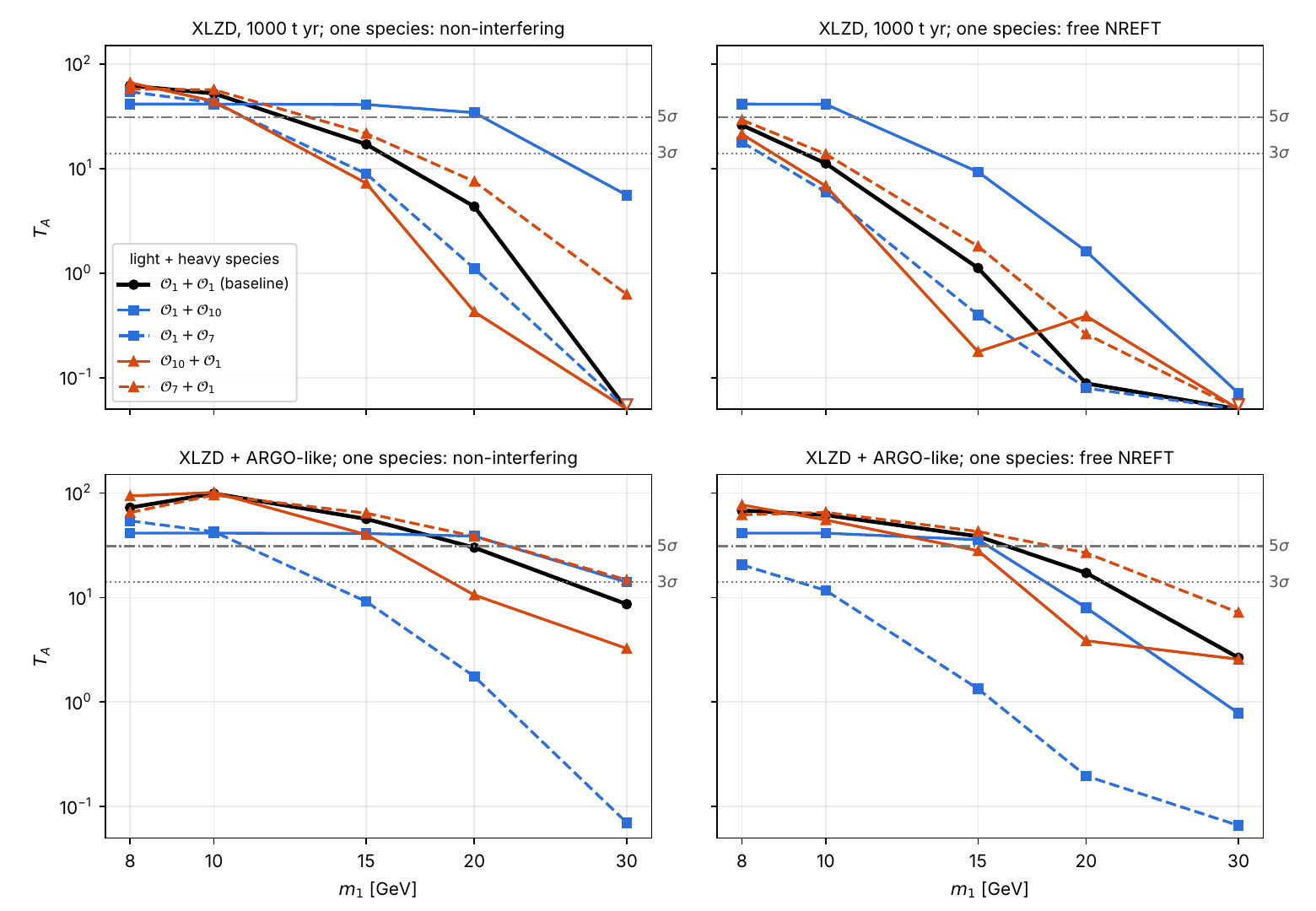}
\caption{Expected test statistic $T_A$ as a function of the mass of the lighter species, for $m_2=350$~GeV, $w=0.5$ and two species with the operators indicated (lighter species first). Top: XLZD alone; bottom: XLZD and an ARGO-like experiment. Left: non-interfering one-component hypothesis; right: free NREFT. Horizontal lines show the $3\sigma$ and $5\sigma$ thresholds. Open triangles on the lower edge mark values of $T_A$ below the plotted range.}
\label{fig:mixed}
\end{figure}

So far both species have had the standard interaction. We now let each species interact through a single operator, $\Op1$, $\Op7$ or $\Op{10}$, with isoscalar couplings, and consider the four combinations in which one of the species has the standard interaction and the other a spin-dependent one. The signal is again at the maximum allowed by LZ, with $w=0.5$. Table~\ref{tab:mixed} gives the number of points of a reduced grid, $m_1\in\{8,10,15,20,30\}$~GeV and $m_2\in\{50,100,350,1000\}$~GeV with $m_2\ge1.8\,m_1$, at which the one-component hypotheses are rejected, and figure~\ref{fig:mixed} shows $T_A$ as a function of $m_1$ for $m_2=350$~GeV. The baseline of two $\Op1$ species is included for comparison.

\begin{table}[t]
\centering
\begin{tabular}{@{}lccc@{\hspace{1.5em}}ccc@{}}
\toprule
 & \multicolumn{3}{c}{XLZD, $1000\tyr$} & \multicolumn{3}{c}{XLZD + ARGO-like} \\
light + heavy & $\Op1$ & non-int. & free & $\Op1$ & non-int. & free \\
\midrule
$\Op1+\Op1$ & 8 [11] & 8 [11] & 0 [4] & 13 [16] & 12 [15] & 9 [13] \\
$\Op1+\Op{10}$ & 15 [19] & 14 [15] & 7 [8] & 17 [19] & 14 [17] & 10 [11] \\
$\Op1+\Op7$ & 8 [8] & 8 [8] & 0 [3] & 8 [14] & 8 [8] & 0 [3] \\
$\Op{10}+\Op1$ & 8 [8] & 8 [8] & 0 [3] & 10 [16] & 10 [11] & 8 [10] \\
$\Op7+\Op1$ & 8 [12] & 8 [11] & 0 [5] & 13 [17] & 13 [17] & 9 [13] \\
\bottomrule
\end{tabular}
\caption{Number of points of the reduced grid, out of 19, at which the median experiment rejects the one-component hypothesis at $5\sigma$ [$3\sigma$], for two species with the operators indicated (lighter species first) and $w=0.5$. When the species have different interactions, rejecting the $\Op1$ hypothesis tests the interaction rather than the number of species, and only the non-interfering and free-NREFT columns are relevant.}
\label{tab:mixed}
\end{table}

Against a single $\Op1$ species, part of the significance comes from the interaction rather than the number of species: for $\Op1+\Op{10}$, for instance, this hypothesis is rejected at $3\sigma$ even for $m_1\simeq30$~GeV. As explained in section~\ref{sec:framework}, rejecting it only shows that the interaction is not $\Op1$ alone. The relevant comparison is with the non-interfering and free-NREFT hypotheses, which contain all three operators.

The most favorable case is a heavy species with $\Op{10}$. Its spectrum, enhanced by $q^2$, extends almost flat to high recoil energies, and no single species can produce both this tail and the steep spectrum of the light species: a light species cannot reach high energies, and a heavy one cannot reproduce the low-energy part. In xenon alone, the non-interfering hypothesis is then rejected at $5\sigma$ up to $m_1\simeq20$~GeV for $m_2\gtrsim350$~GeV (at $(20,350)$~GeV, $T_A\simeq34$, compared with $4.3$ for two $\Op1$ species), and even the free NREFT is rejected at $5\sigma$ for $m_1\lesssim10$~GeV and $m_2\gtrsim100$~GeV.

The least favorable case is a heavy species with $\Op7$. Its velocity-suppressed spectrum is easier to imitate, and in xenon alone the non-interfering hypothesis is rejected at $3\sigma$ only for $m_1\lesssim10$~GeV. Argon does not help: the heavy species produces no events in argon, whose nucleus has no spin, and the single-species fit, which explains the xenon tail with $\Op7$, predicts none either.

When the lighter species is the one with a spin-dependent interaction, the results are close to those of the baseline. A light $\Op7$ species is slightly easier to identify, and a light $\Op{10}$ species somewhat harder; with ARGO, which observes the heavy $\Op1$ species, the window extends as in the baseline for a light $\Op7$ species and somewhat less for a light $\Op{10}$ species.

In all cases the window remains confined to light species. For $m_1\simeq30$~GeV, no combination reaches $5\sigma$ against the non-interfering hypothesis, even with ARGO, and the free NREFT is never rejected at $3\sigma$. The conclusions of the previous sections therefore do not rely on the assumption that both species have the standard interaction.

\section{Discussion}
\label{sec:discussion}
\subsection{Interactions and the number of species}
Our results separate two effects. The first, the event budget together with kinematics, limits the identification of two species even against a single species with the standard interaction, and confines it to light species. The second, the freedom of the interaction of the single species, matters only within that range, and only if interference is allowed: operators that can only add events to the $\Op1$ spectrum barely reduce the significance, and with a large argon exposure hardly at all, while the $\Op1$--$\Op3$ interference removes most of it in xenon, at the price of a strong cancellation and a large coefficient for the most suppressed operator. A claim that DM consists of more than one species therefore depends on whether such interference is allowed in the single-species alternative. When the two species have different interactions, that alternative must contain the operators of both, and only its rejection, not that of a single $\Op1$ species, establishes that there is more than one species. The degeneracy between the mass and the interaction of a single DM species found in refs.~\cite{Gluscevic:2015sqa,Edwards:2018lsl,Krauss:2018} thus reappears here as a degeneracy between the number of species and their interactions; likewise, identifying specific properties of the interaction, such as P or CP violation, requires tens to hundreds of signal events~\cite{Catena:2020tbv}, comparable to the whole event budget allowed by LZ.

Our results agree with the kinematic part of the picture of refs.~\cite{Profumo:2009tb,HerreroGarcia:2017vrl,HerreroGarcia:2018qnz}, in which two species can be identified for sufficiently different masses, but show that, once the signal is capped by current limits, the exposures are those now planned and backgrounds are included, the window shrinks to light species.

\subsection{Dependence on the halo}
The most important assumption beyond those of the effective theory concerns the halo. We have assumed that both species share a known velocity distribution and varied its parameters only within the Standard Halo Model. This assumption is stronger than it may seem. In a single target and for $\Op1$, the rate of a single species is proportional to $F^2(E_R)\,\eta(v_{\min}(E_R))$, where $F$ is the nuclear form factor and $\eta$ a non-increasing function of $v_{\min}$. For two species, the rate is proportional to $F^2(E_R)\,[a_1\eta(v_{\min,1})+a_2\eta(v_{\min,2})]$, with $a_i\ge0$ and $v_{\min,i}=\sqrt{m_TE_R/2\mu_i^2}$. For any trial mass $m_\star$, with reduced mass $\mu_\star$ and $u=\sqrt{m_TE_R/2\mu_\star^2}$, we have $v_{\min,i}=(\mu_\star/\mu_i)\,u$, so the rate of two species equals that of a single species of mass $m_\star$ with
\begin{equation}
\eta_{\rm eff}(u)=a_1\,\eta\Big(\frac{\mu_\star}{\mu_1}\,u\Big)+a_2\,\eta\Big(\frac{\mu_\star}{\mu_2}\,u\Big),
\end{equation}
which is non-increasing and therefore corresponds, up to a normalization absorbed in the cross section, to some velocity distribution. Up to the small differences between isotopes, which have different reduced masses, a single target with an unknown velocity distribution cannot distinguish one species from two~\cite{Fox:2010vz,Fox:2010bz,Gondolo:2012rs,DelNobile:2013cva,Feldstein:2014gza}. Identifying two species in a single target therefore always depends on the halo model. The degeneracy also works in the opposite direction: a feature in the velocity distribution, such as a stream, adds a step to $\eta$ and produces a change of slope in the spectrum of a single species, which an analysis that assumes the Standard Halo Model could mistake for a second species. Two targets, which probe different ranges of $v_{\min}$ at a given energy, can reduce this dependence if the species share a velocity distribution. This is a further reason to combine xenon and argon, although a halo-independent analysis of the combination is beyond the scope of this work.

\subsection{Other targets and simplifications}
Other targets could provide further information. The single-species imitations of section~\ref{sec:shape} that rely on $\Op7$ and $\Op{10}$ predict events in fluorine, whose nucleus has a large spin, but current bubble chamber limits do not exclude them (appendix~\ref{app:pico}). Annual modulation, which differs between velocity-suppressed operators and $\Op1$~\cite{Witte:2016ydv}, could also help, but requires exposures beyond those considered here.

Several simplifications affect the details of our results. We work with the recoil energy rather than the measured scintillation and ionization signals, and the event budget treats the LZ limit as a fixed number of events, an approximation that we estimated in section~\ref{sec:budget} to move the boundaries by about $1$~GeV. The low-mass window depends on how we model the threshold and the $^8$B background, and the shell-model spin responses enter the imitations that use $\Op7$ and $\Op{10}$. Our results apply to WIMP-like DM: our mass grid extends from $8$~GeV to $1.5$~TeV, and xenon loses sensitivity to nuclear recoils below about $6$~GeV, so lighter DM, and in particular sub-GeV DM, would require a separate analysis.

We have used recoil energies below $60\keVnr$. For two $\Op1$ species and their best single-species fits, less than $0.5\%$ of the events lie between $60$ and $120\keVnr$, so a wider energy range would add little. A heavy species with $\Op{10}$ ($\Op3$), instead, has about $0.8$ ($0.4$) times as many events between $60$ and $120\keVnr$ as below $60\keVnr$, so our results for two species with $\Op1$ and $\Op{10}$ are conservative. Such a species is also constrained by the LZ searches extended to about $270\keVnr$~\cite{LZ:2023lcz,LZ:2026axp}: at our cap, it would produce about two events above $60\keVnr$ in the $2.84\tyr$ of ref.~\cite{LZ:2026axp}, where the expected background is far below one event and one event was observed. This comparison of event counts suggests a constraint comparable to our cap, but establishing it would require a recast of that search with the full spectrum and backgrounds. In the mixed benchmarks of section~\ref{sec:mixed}, with $w=0.5$, the heavy species would produce only about one such event.

Finally, we have assumed elastic contact interactions with isoscalar couplings, and restricted the alternative to the four operators of the standard basis common to all DM spins. Isospin-violating couplings, higher-order operators, or operators involving the DM spin would give the alternative more freedom and could only reduce the window further (section~\ref{sec:formalism}). Inelastic scattering or light mediators, instead, would change the spectra of both the signal and the alternative, and could widen or narrow the window; they require a separate analysis. The calibration of the test statistic is local, using the best one-component fit as the null hypothesis, and the $5\sigma$ threshold extrapolates the tail with a $\chi^2$ fit. None of these limitations is likely to change the main picture, which is controlled by the event budget and by kinematics, but they affect the precise position of the contours.

\section{Conclusions}
\label{sec:conclusions}
Determining the nature of DM remains one of the most important problems in particle and astroparticle physics. A basic part of that problem is how many species DM consists of: just one, as is usually assumed, or several, as in multi-component DM scenarios. Gravity, which provides all the evidence for DM so far, cannot measure this number; it must be determined by experiments that respond to individual DM particles.

Whether direct detection experiments can establish that DM consists of more than one species is an old question, but it can now be answered for the experiments that are planned and for the signal that current limits still allow. We have addressed it for WIMP-like DM and the xenon and argon detectors XLZD, DarkSide-20k and ARGO. If a signal is observed, establishing that it comes from more than one species requires excluding not only a single species with the standard spin-independent interaction ($\Op1$), but also a single species with a different interaction. We have therefore compared two species, the simplest multi-component case, both with $\Op1$ or with different interactions, against a single species described by the four operators of the non-relativistic effective theory that are common to all DM spins. We have set the signal at the maximum allowed by LZ, included the neutrino and electron-recoil backgrounds, and calibrated the test statistic with pseudo-experiments.

The main limitation is the small number of events that current limits still allow. Even assuming the standard spin-independent interaction, XLZD, with its largest envisaged exposure of $1000\tyr$, can reject a single species at an estimated $5\sigma$, for the Standard Halo Model, only if the lightest species is below about $12$~GeV ($15$~GeV at $3\sigma$) and both contribute comparably to the signal. With its baseline exposure of $200\tyr$, XLZD reaches at most about $3\sigma$, for $m_1\lesssim10$~GeV, and never $5\sigma$. Kinematics sets this limit: as the lighter species becomes heavier, its spectrum approaches that of the heavier one, and the few hundred events allowed can no longer tell them apart. Below about $12$~GeV, operators that do not interfere with $\Op1$ barely reduce the significance, while a free combination of the four operators removes the $5\sigma$ reach in xenon, through a strong cancellation between $\Op1$ and $\Op3$. DarkSide-20k adds little, but an argon experiment with the exposure of ARGO extends this reach to about $15$--$20$~GeV at $5\sigma$ ($20$--$25$~GeV at $3\sigma$) and makes it much less sensitive to the assumed interaction and to electron-recoil leakage. If the two species have different interactions, the reach can be larger, in particular when the heavier one has a momentum-dependent interaction, but at $5\sigma$ it never extends to $m_1\simeq30$~GeV.

Because XLZD and ARGO are the ultimate detectors of this kind, and because the limitation comes from kinematics and from the event budget fixed by current limits rather than from the particular detectors, direct detection is unlikely ever to tell, even at $3\sigma$, whether DM consists of one species or several when all of them are heavier than about $30$~GeV, within the assumptions of this work. In practice, the first test will come from XLZD at its baseline exposure, possibly together with DarkSide-20k, which is the configuration with the least reach, while ARGO, which makes the difference, is the most distant in time.

Establishing in direct detection that DM consists of more than one species is thus possible, but only if the lightest species is below about $20$~GeV, where ARGO can reach $5\sigma$, with two targets to reduce the dependence on the halo model, and with explicit assumptions about the interactions allowed for a single species.

\appendix
\section{Calibration of the test statistic}
\label{app:calib}
We calibrate mainly the non-interfering hypothesis, in which the fit is linear in non-negative amplitudes, and check the other two one-component hypotheses at one benchmark each. For each benchmark we generate at least $150$ pseudo-data sets from the best one-component fit to the Asimov data set (null) and $150$ from the two-component benchmark (alternative), and fit both hypotheses to each of them, with the masses determined on a grid and refined continuously. We checked the fitter against the one used for the Asimov maps; the two agree to $0.01$ in $T$. For the $(15,100)$~GeV benchmark in xenon we extended the null sample to $3000$ pseudo-experiments, which we use in table~\ref{tab:calib}.

All fits minimize $-2\ln\mathcal L$ with the L-BFGS-B algorithm and analytic gradients, with the background normalizations constrained by $\theta_j\ge0$. For the Asimov maps, the mass of the single species is scanned on $75$ logarithmically spaced values between $6$~GeV and $30$~TeV and the minimum refined continuously in $\ln m$; at each mass the coefficients are started from each operator alone and from random directions, and for the free NREFT from both relative signs of $c_1$ and $c_3$. In the pseudo-experiments, the mass is scanned on $34$ values between $6$~GeV and $3$~TeV and the two masses of $H_2$ on $16\times16$ values, followed by a continuous refinement. Values $T\le0$ are counted as $T=0$, and the $p$-value of an observed $T$ is the probability, under the fitted null distribution, of a value at least as large.

\begin{figure}[t]
\centering
\includegraphics[width=\textwidth]{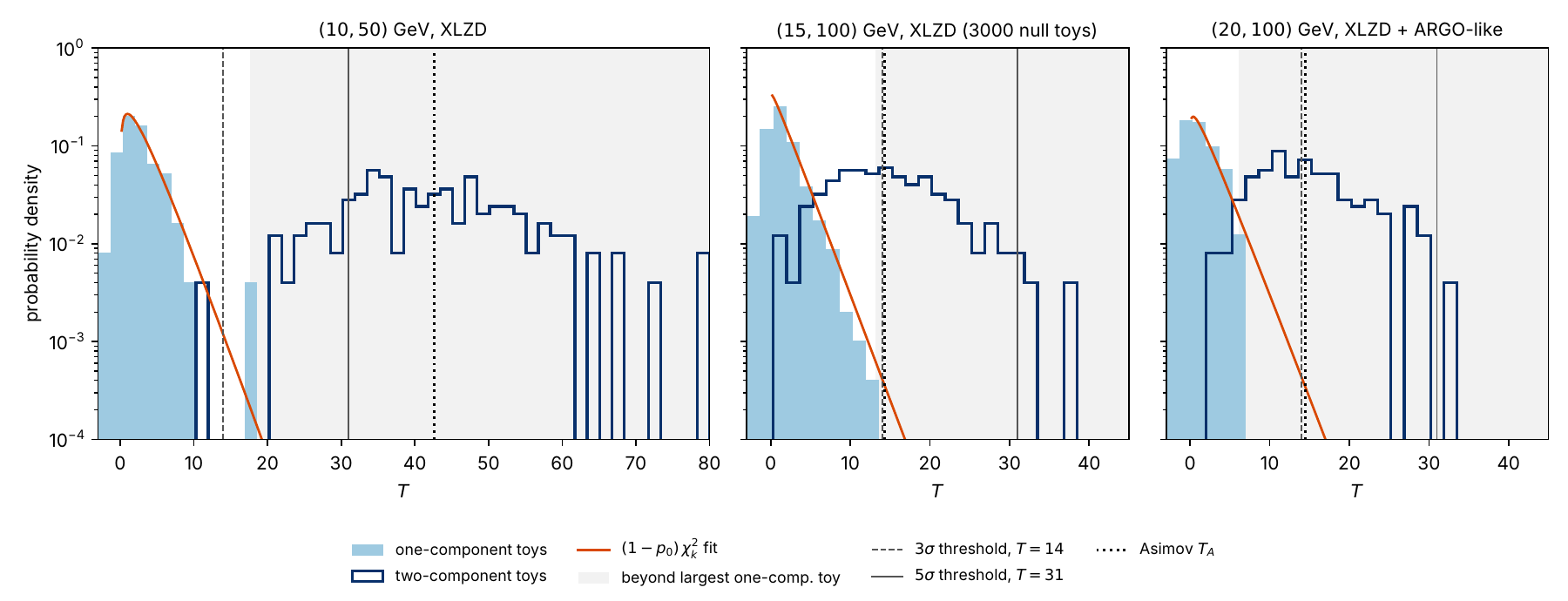}
\caption{Distribution of $T$ for pseudo-experiments generated under the best one-component fit (filled) and under the two-component benchmark (line), with the fit to the one-component distribution and the Asimov value $T_A$, for the non-interfering hypothesis and $w=0.5$. The vertical lines show the $3\sigma$ and $5\sigma$ thresholds, $T=14$ and $31$; in the shaded region, beyond the largest one-component pseudo-experiment, the distribution is known only through the fitted tail. The central panel uses $3000$ one-component pseudo-experiments.}
\label{fig:calib}
\end{figure}

Figure~\ref{fig:calib} and table~\ref{tab:calib} show the results. The null distribution has a fraction $p_0$ of non-positive values, which occur when the two-component fit does not improve on the one-component one, and a tail that we fit with $(1-p_0)\chi^2_k$. We obtain the $3\sigma$ and $5\sigma$ thresholds from this fit. For the non-interfering hypothesis, the median of the alternative distribution agrees with $T_A$ in all cases. For a single $\Op1$ species the thresholds are the same, and for the free NREFT, for which only $21\%$ of the null pseudo-experiments give $T>0$, so that the fitted tail is more uncertain, they are slightly lower; for both, the median of the alternative distribution is about $11\%$ below $T_A$.

\begin{table}[t]
\centering
\small
\setlength{\tabcolsep}{3pt}
\begin{tabular}{@{}llcccccc@{}}
\toprule
benchmark & $H_1$ & $T_A$ & $p_0$ & $k$ & $T_{3\sigma}$ / $T_{5\sigma}$ & power at $3\sigma$ / $5\sigma$ & median $T$ \\
\midrule
$(10,50)$, XLZD & non-int. & 42.7 & 0.13 & 3.0 & 15.3 / 32.9 & 0.99 / 0.79 & 41.6 \\
$(15,100)$, XLZD$^\ast$ & non-int. & 14.3 & 0.20 & 2.1 & 13.0 / 30.0 & 0.58 / 0.03 & 14.0 \\
$(20,100)$, XLZD + ARGO-like & non-int. & 14.5 & 0.39 & 2.4 & 13.1 / 30.3 & 0.54 / 0.01 & 14.1 \\
$(15,100)$, XLZD & $\Op1$ & 15.2 & 0.23 & 2.5 & 14.0 / 31.3 & 0.47 / 0.02 & 13.5 \\
$(8,350)$, XLZD & free & 26.0 & 0.79 & 2.7 & 11.6 / 29.1 & 0.95 / 0.19 & 23.0 \\
\bottomrule
\end{tabular}
\caption{Calibration of the test statistic for five benchmarks and one-component hypotheses $H_1$ (non-interfering, $\Op1$ only, free NREFT), with $w=0.5$, xenon $1000\tyr$ and, where indicated, argon $3000\tyr$. $p_0$ is the fraction of non-positive values under the one-component hypothesis and $k$ the number of degrees of freedom of the fitted $\chi^2$ tail; the power is the fraction of two-component pseudo-experiments above the threshold, and the last column the median of $T$ under the two-component hypothesis. $^\ast$Null distribution from $3000$ pseudo-experiments; for the other benchmarks, $150$.}
\label{tab:calib}
\end{table}

\section{Bubble chamber limits on single-species imitations}
\label{app:pico}
The imitations of section~\ref{sec:shape} that rely on $\Op7$ and $\Op{10}$ have large spin-dependent couplings, which could be constrained by fluorine targets. We model PICO-60~\cite{PICO:2019vsc} with its two exposures of $C_3F_8$, $1167$~kg\,day at a threshold of $3.3$~keV and $1404$~kg\,day at $2.45$~keV, and a nucleation efficiency that rises linearly from the threshold to twice its value. Calibrating, as for LZ, with the published limit on the spin-dependent proton cross section at $25$~GeV, we find that the limit corresponds to $6.4$ events. Each free-NREFT imitation of table~\ref{tab:shape}, normalized to the maximal xenon signal, predicts at most about $0.1$ events in PICO-60, while the two-component signal predicts at most about $10^{-3}$. This is an order-of-magnitude estimate, but the margin is large: bubble chamber limits do not exclude these imitations. A future fluorine exposure of order $10^2$--$10^3$~t\,day, depending on the masses, would test them for isoscalar couplings.


\end{document}